\documentclass[aps,prl,reprint,superscriptaddress,10pt]{revtex4-2}

\usepackage{graphicx,amsmath,amssymb,dcolumn,physics,color,wasysym}

\graphicspath{{./figs/}{./}}

\begin{document}

\title{Resolving the binding-kinase discrepancy in bacterial chemotaxis: A nonequilibrium allosteric model and the role of energy dissipation}

\author{David Hathcock}
\email{These two authors contributed equally}
\affiliation{IBM T.~J.~Watson Research Center, Yorktown Heights, NY 10598}

\author{Qiwei Yu}
\email{These two authors contributed equally}
\affiliation{IBM T.~J.~Watson Research Center, Yorktown Heights, NY 10598}
\affiliation{Lewis-Sigler Institute for Integrative Genomics, Princeton University, Princeton, NJ 08544}

\author{Bernardo A.~Mello}
\affiliation{International Center of Physics, Physics Institute, University of Brasilia, Brasilia, Brazil}

\author{Divya N.~Amin}
\affiliation{Department of Biochemistry, University of Missouri, Columbia, MO 65211}

\author{Gerald L.~Hazelbauer}
\affiliation{Department of Biochemistry, University of Missouri, Columbia, MO 65211}

\author{Yuhai Tu}
\affiliation{IBM T.~J.~Watson Research Center, Yorktown Heights, NY 10598}

\date{\today}

\begin{abstract}
    \textbf{Abstract:} The \textit{Escherichia coli} chemotaxis signaling pathway has served as a model system for studying the adaptive sensing of environmental signals by large protein complexes. The chemoreceptors control the kinase activity of CheA in response to the extracellular ligand concentration and adapt across a wide concentration range by undergoing methylation and demethylation.
    Methylation shifts the kinase response curve by orders of magnitude in ligand concentration while incurring a much smaller change in the ligand binding curve.
    Here, we show that this asymmetric shift in binding and kinase response is inconsistent with equilibrium allosteric models regardless of parameter choices.
    To resolve this inconsistency, we present a nonequilibrium allosteric model that explicitly includes the dissipative reaction cycles driven by ATP hydrolysis.
    The model successfully explains all existing measurements for both aspartate and serine receptors. Our results suggest that while ligand binding controls the equilibrium balance between the ON and OFF states of the kinase, receptor methylation modulates the kinetic properties (e.g., the phosphorylation rate) of the ON state.
    Furthermore, sufficient energy dissipation is necessary for maintaining and enhancing the sensitivity range and amplitude of the kinase response.
    We demonstrate that the nonequilibrium allosteric model is broadly applicable to other sensor-kinase systems by successfully fitting previously unexplained data from the DosP bacterial oxygen-sensing system.
    Overall, this work provides a new perspective on cooperative sensing by large protein complexes and opens up new research directions for understanding their microscopic mechanisms through simultaneous measurements and modeling of ligand binding and downstream responses.

    \textbf{Significance:} Adaptation in bacterial chemotaxis is carried out by receptor methylation, which shifts the kinase response curve by orders of magnitude in ligand concentration. However, the receptor-ligand binding curve is only shifted slightly by receptor methylation. We show that this discrepancy in shifts rules out the previously proposed equilibrium allosteric models for the system. We develop a nonequilibrium model that  explicitly includes the phosphorylation-dephosphorylation cycle driven by ATP hydrolysis. Our model agrees with all the existing experiments and shows that while ligand binding affects the ON/OFF switching of the kinase activity, receptor methylation affects the ON-state kinetics. Our study also shows that a minimum energy dissipation is needed to maintain the observed sensitivity range and amplitude of the kinase response.
\end{abstract}

\maketitle

\section{Introduction}

Most biological machines that are responsible for important functions are made of multiple components (proteins and RNAs) that work together in a cooperative manner. Examples include the ribosome for protein synthesis~\cite{ramakrishnan_ribosome_2002} and the bacterial flagellar motor for locomotion~\cite{berg_rotary_2003}. One of the most important models in describing the cooperative function of a large protein complex with multiple subunits is the Monod-Wyman-Changeux (MWC) model~\cite{MWC1965}. Originally developed to describe cooperative allosteric interactions in a multi-subunit enzyme such as hemoglobin~\cite{MWC1965,Edelstein1971Extensions}, it has been extended to describe signal transduction~\cite{Changeux05} in large protein complexes such as the bacterial chemoreceptor cluster with heterogeneous components~\cite{victor2004interaction,Mello05,mwcwingreen} and gene regulation by transcription regulators~\cite{RAZOMEJIA2018Tuning} (see the recent book by Phillips~\cite{Phillips2020MolecularSwitch} for a comprehensive review). However, despite its many successes, the MWC model is highly simplified, assuming equilibrium interactions between components of the protein complex and a two-state (all-or-none) behavior for the entire complex.
We now know that many biological machines operate out of equilibrium through the hydrolysis of energy-rich molecules such as nucleotide triphosphate (NTP) including ATP and GTP.
Moreover, rich structural information of these complexes has been revealed by high-resolution imaging techniques such as cryo-electron tomography~\cite{sriram2007direct,liu_molecular_2012}.
In light of this new information, it becomes necessary to examine the validity of the MWC assumptions and to elucidate whether the MWC model still provides a faithful description of the underlying process in these protein complexes.

Here, we reexamine the applicability of the MWC model to signal transduction in chemoreceptor clusters found in almost all bacteria~\cite{briegel2009universal}.
Bacteria use these membrane-bound chemoreceptors to sense and to respond to changes in their environments, such as chemical concentrations, temperature, pH, and osmotic pressure~\cite{hazelbauer2008bacterial}. There are around 20,000 methyl-accepting chemotaxis proteins (MCP) in an \emph{Escherichia coli} cell~\cite{Li04}.
They form large clusters near the cell poles~\cite{Maddock93} and serve important cellular functions such as signal amplification~\cite{Bray98, Sourjik02, Mello03a} and enhancing adaptation~\cite{pontius_adaptation_2013,wingreen2006neighborhoods,li_adaptational_2005}.
Together with quantitative functional experiments, modeling work using variations of the MWC model has played an important role in understanding the mechanisms underlying key functions such as signal integration, adaptation, and amplification~\cite{tu2013quantitative, sourjik_responding_2012}.

The \textit{E. coli} chemotaxis signaling pathway involves receptor complexes composed of membrane-bound chemoreceptors, cytoplasmic histidine kinase CheA, and an adaptor protein CheW.
The complex transduces the signal induced by the binding of ligands to the chemoreceptors to the kinase activity of CheA, which modulates the swimming behavior of the bacterium by phosphorylating the intracellular response regulator CheY.
In order to remain sensitive at varying levels of external stimuli, the kinase activity is modulated by an adaptation mechanism.
Adaptation is achieved by the methylation and demethylation of the chemoreceptor catalyzed by CheR and CheB, respectively.

The kinase response of the complex can be successfully described by a generalized MWC model~\cite{Mello05}, which captures the significant change in the sensitivity of kinase response as receptor methylation level changes. Since the MWC model is an equilibrium model that satisfies detailed balance in all transitions between internal states, it predicts that varying the receptor methylation level should induce similar changes in both ligand occupancy and kinase response~\cite{Mello07}. However, this prediction is inconsistent with \textit{in vitro} experiments by Borkovitch et al.~\cite{Borkovich92} and Amin and Hazelbauer~\cite{Amin2010Chemoreceptors} for aspartate receptors (Tar) and by Levit and Stock~\cite{Levit02} for serine receptors (Tsr), which showed that while changes in receptor methylation shift the kinase response curve over a significant range of ligand concentration, the corresponding shift in ligand occupancy is much smaller.
Moreover, Vaknin and Berg~\cite{Vaknin07} measured the receptor response \emph{in vivo} in the absence of both the histidine kinase CheA and linker protein CheW.
As the methylation level increases, they again found a much smaller shift in the response of bare receptor oligomers than that of the kinase response of the full complex.
So far, this fundamental discrepancy remains unexplained.

In this paper, we present a nonequilibrium model that explains both the ligand binding and the kinase activity for different receptor methylation levels.
First, we develop a parametric test, which systematically demonstrates that the existing measurements cannot be consistently explained by the MWC model or similar equilibrium models. This motivates us to develop a new nonequilibrium model, which extends the MWC approach by adding a kinetic description of the ATP-driven phosphorylation-dephosphorylation(PdP) cycle of the downstream signaling molecule CheY controlled by the kinase CheA and the phosphatase CheZ.
The model is capable of consistently fitting all available measurements of ligand binding, receptor conformation, and kinase activity for \emph{E. coli} chemotaxis~\cite{Amin2010Chemoreceptors,Vaknin07,Levit02}.
Crucially, the experimentally observed behavior is only enabled by a sufficiently strong nonequilibrium driving in the PdP cycle, which is provided by ATP hydrolysis. If the driving is below certain thresholds, the model fails to simultaneously capture ligand binding and kinase activity, especially the discrepancy in their shifts when receptor methylation level changes.
Finally, this nonequilibrium allosteric model should be generally applicable to other signaling pathways involving dissipation, in particular, sensor-kinase systems that are driven out of equilibrium by the PdP cycle. Indeed, the model successfully captures both the binding and kinase activity measurements in the bacterial oxygen-sensing system DosP~\cite{tuckerman_oxygen-sensing_2009}.

\section{Results}

\subsection{Equilibrium allosteric models fail to explain both ligand binding and kinase response}

\begin{figure*}[t]
    \includegraphics[width=.65\linewidth]{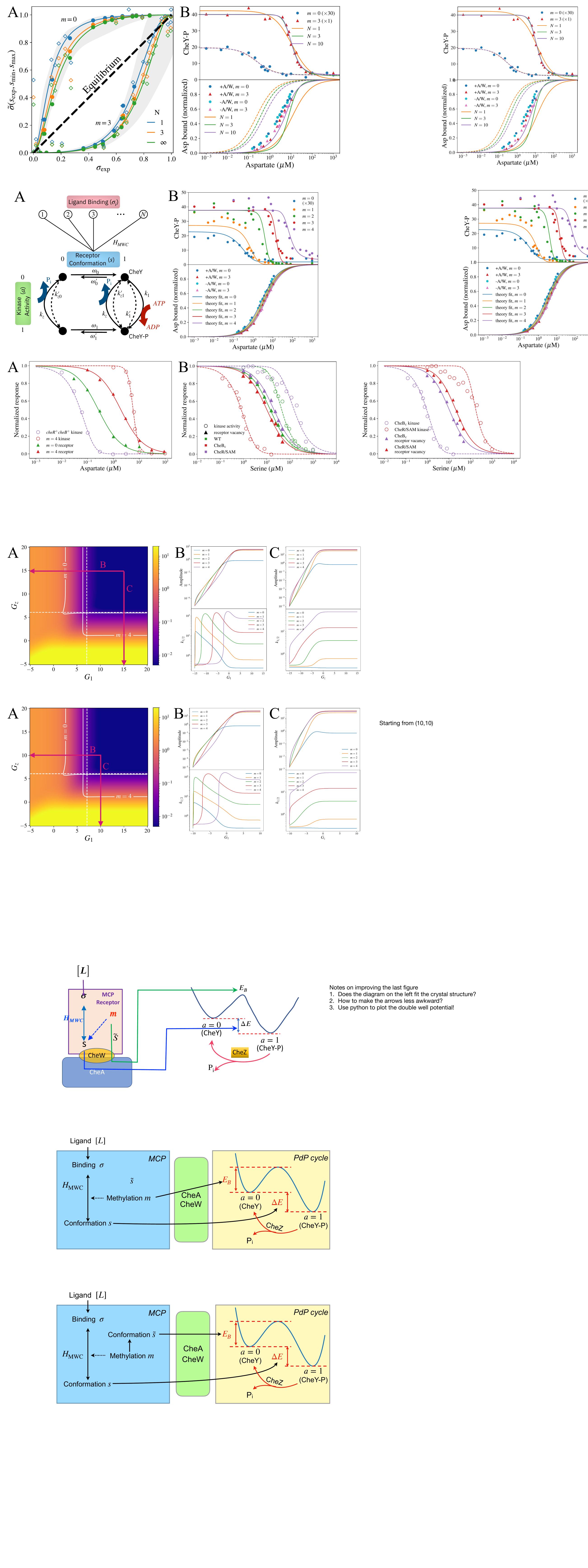}
    \caption{\label{equilibirum_test}
        Measured aspartate binding and kinase activity of Tar receptor signaling complexes~\cite{Amin2010Chemoreceptors} are inconsistent with equilibrium MWC models.
        (A) A parametric test plotting the measured aspartate receptor occupancy ($\sigma_\mathrm{exp}$) against the inferred occupancy  $\tilde \sigma(s_{\rm{exp}}, s_{\max}, s_{\min})$ from MWC model with $N=1, 3$, and $\infty$ and for methylation levels $m=0$ and $3$. The solid curves correspond to independent Hill function fits to the data,
        the filled circles ($\newmoon$) are the $(\sigma_\mathrm{exp},\tilde{\sigma})$ pairs inferred from binding measurements, and the open diamonds ($\Diamond$) are inferred from measurements of CheY phosphorylation.
        The gray shaded area shows 95\% confidence regions from the Hill function fits. The transformed data lie far from the diagonal (black dashed line), indicating that the system is nonequilibrium.
        (B) Fitting the kinase activity for $m=0$ (dashed) and $m=3$ (solid) with equilibrium MWC models ($N=1,3,10$) captures the CheY-P measurements (upper) but not the aspartate binding (lower).
    }
\end{figure*}

To start, we focus on the \textit{in vitro} measurements by Amin and Hazelbauer on Tar receptors embedded in native membrane vesicles~\cite{Amin2010Chemoreceptors}.
The receptors were fixed at different levels of methylation by substituting glutamates (E) with glutamine (Q) at the receptor methyl-accepting sites.
By comparing dose-response curves of receptors with zero or three modifications, it was found that methylation shifts the kinase response curve significantly but only induces a much smaller shift in ligand binding (see Fig.~\ref{equilibirum_test}B).

Before introducing the nonequilibrium allosteric model, we first examine whether the behaviors (especially ligand binding and kinase activity) of the chemoreceptors can be consistently described by an equilibrium allosteric model.
To this end, we present a parametric test that systematically detects inconsistency between the measurements and the MWC model without fitting the data to any specific function.
Other equilibrium models, such as Ising models, can be ruled out using a similar approach (see SI Appendix).

In the MWC model, the receptor complex has $N$ identical binding sites, whose occupancy is denoted by $\sigma_i$ $(i = 1, 2 \dots N)$; $\sigma_i=0, 1$ represents a vacant or occupied receptor, respectively. The receptor methylation level is denoted by $m$.
The MWC model assumes all-or-none behavior for the whole complex. Namely, the receptor activity $s$ is either on ($s=1$) or off ($s=0$).
The energy (Hamiltonian) of the receptor complex is given by
\begin{equation}\label{MWCenergy}
    H = (-\mu + E_0 s) \sum_{i=1}^N \sigma_i + E_s s,
\end{equation}
with $\mu = \log\qty([L]/K_i)$ being the chemical potential for binding, which depends on ligand concentration $[L]$ and a (methylation-dependent) dissociation constant $K_i(m)$. $E_s$ is the energy difference between the active and inactive receptor states in the absence of a ligand. Each occupied binding site increases this energy difference by $E_0>0$, thereby suppressing activity at high occupancy.

Given the MWC Hamiltonian, the average receptor activity is
\begin{equation}\label{avActivity}
    \expval{s} = \left ( 1 +e^{E_s} \qty( \frac{1+ [L]/K_i}{1+ e^{-E_0} [L]/K_i})^N\right)^{-1},
\end{equation}
and the average binding is
\begin{equation}\label{avBinding}
    \expval{\sigma} =\frac{[L]}{K_i+ [L]} (1 -\langle s \rangle) + \frac{ [L]}{e^{E_0}K_i+ [L]} \langle s \rangle.
\end{equation}
In the limit of large $E_s$, we have  $\expval{s} \ll 1$, so the binding curve becomes a Hill function with Hill coefficient $n_H=1$ (no cooperative binding), consistent with experiments using vesicle-bound chemoreceptors~\cite{Amin2010Chemoreceptors}.
The maximum activity $s_{\max} = \qty(1+\exp(E_s))^{-1}$ is reached at $[L]=0$, and the minimum activity $s_{\min} = \qty(1+\exp(E_s + N E_0))^{-1}$ occurs at $[L]=\infty$.

The MWC model has seen great success in capturing the methylation-dependence of the downstream CheY-P response~\cite{Mello05, Mello07}, even in the presence of time-varying stimuli~\cite{shimizu2010modular}. These studies suggest the following basic mechanism: methylation affects the energy for the active state, $E_s(m)$, and thereby shifts the kinase response curves across orders of magnitude in ligand concentration. However, this mechanism does not take into account measurements of ligand binding.

Here, we examine whether the binding and activity curves measured at the same methylation level are both compatible with the equilibrium MWC model. This can be done by eliminating the concentration variable $[L]$ from Eqs.~(\ref{avActivity}) and (\ref{avBinding}) to obtain a parametric relation between $\expval{s}$ and $\expval{\sigma}$.
For example, solving Eq.~(\ref{avActivity}) for $[L]$ and substituting the solution into Eq.~(\ref{avBinding}) gives the mean occupancy as a function of the receptor activity,
\begin{equation}\label{equilTransform}
    \expval{\sigma} = \tilde\sigma\qty( \expval{s}, s_{\min}, s_{\max}),
\end{equation}
where $s_{\max}$ and $s_{\min}$ are the maximum and minimum activity mentioned above, which can be determined from experimental measurements of kinase activity. For any fixed $N$, $s_{\max}$ and $s_{\min}$ uniquely determine the value of the energy parameters $E_0$ and $E_s$, which gives $\tilde \sigma$ as a function of $(\expval{s}, s_{\max}, s_{\min})$. The expression of $\tilde\sigma$ for finite $N$ can be complicated, but its behavior can be illustrated in small and large $N$ limits.
For $N=1$, the equilibrium model simplifies to $\tilde \sigma = (s_{\max} - \langle s \rangle)/(s_{\max} - s_{\min})$, which equates ligand binding with the normalized kinase activity.
On the other hand, for large $N$, the inferred receptor occupancy converges to
\begin{equation}
    \tilde \sigma = \log \frac{ s_{\max} (1-\langle s\rangle)}{\langle s\rangle (1-s_{\max})} \bigg/ \log \frac{ s_{\max} (1-s_{\min})}{s_{\min} (1-s_{\max})}.
\end{equation}
The curves for intermediate $N$ lie between these two extreme limits.

To test the validity of the equilibrium MWC model, we transform the measured kinase response curves for two different methylation levels~\cite{Amin2010Chemoreceptors} according to Eq.~(\ref{equilTransform}) and plot the inferred occupancy ($\tilde\sigma$) against the corresponding measurements of the receptor occupancy ($\sigma_\mathrm{exp}$)~\footnote{Following previous studies, we assume the measured CheY-P concentration is proportional to activity in the model: $[\text{CheY-P}] = A_0 \langle s \rangle$, where $A_0$ is the saturating concentration of CheY-P. We use $A_0=60$pM chosen to be slightly above the maximum CheY-P concentration measured in this set of experiments (the results are robust to increasing the value of $A_0$).}. If the equilibrium model is valid, all the points should collapse onto the diagonal $\sigma = \tilde \sigma$ for some choice of $N$.
However, the data for both methylation levels ($m=0,3$) lie well off the diagonal for any choice of $N$ (Fig.~\ref{equilibirum_test}A), indicating that the system (even for a single methylation level) cannot be described by the equilibrium MWC model with any $N$.

The intuition behind this inconsistency is that the MWC model cannot capture the relative shift between binding and response curves.
To demonstrate this, we fit the MWC model only to the measured kinase activity (CheY-P level) and compare the resulting binding and activity curves with experiments (Fig.~\ref{equilibirum_test}B). The MWC model successfully captures kinase activity (upper panel) but does not produce the correct binding curves (lower panel).
Instead, it predicts that the increase in binding and the decrease in kinase activity should occur at around the same ligand concentration. This prediction is inconsistent with experiments, which found the sharpest change in kinase activity occurring at a concentration that is either much lower ($m=0$) or higher ($m=3$) than that of binding. Moreover, as the methylation level changes from $m=0$ to $m=3$, the shift in activity curves is much greater than the shift in binding curves.
These results suggest that the MWC model is unable to capture the data either at a single methylation level or the change between different methylation levels.
Similarly, fitting the MWC model to the measured binding curves results in discrepancies with the measured kinase activity (see SI Appendix).
Note that these fittings are only shown to build intuition.
They are all encompassed by the parametric test (Fig.~\ref{equilibirum_test}A), which offers stronger evidence by revealing inconsistency with the MWC model without relying on fitting with any assumptions or any particular formulation of the loss function.

Similar results hold for more complex equilibrium models: for example, if we treat the measured kinase activity $a$ as a separate degree of freedom from the receptor activity $s$. Assuming the two activities are coupled by equilibrium mechanisms, with the active receptor promoting kinase phosphorylation, we add the following terms to the MWC Hamiltonian,
\begin{equation}\label{kinaseActivityEnergy}
    H_a = (F_0 + \Delta F s ) a,
\end{equation}
where $F_0$ is the energy of the active kinase when the receptor is inactive and $F_1 = F_0 + \Delta F < F_0$ is the energy when the receptor is active. The average activities of the joint Hamiltonian $H+H_a$ are linearly related,
\begin{equation}\label{activityLinearRelation}
    \frac{a_{\max} -\langle a \rangle }{a_{\max} - a_{\min}} = \frac{s_{\max} -\langle s \rangle }{s_{\max} - s_{\min}}.
\end{equation}
Therefore, our analysis above also applies to equilibrium models with additional binary degrees of freedom. In the SI Appendix, we show that this linear relation holds for arbitrary chains of binary variables coupled via equilibrium interactions. Hence, these types of models are also inconsistent with the experiments.

Beyond MWC models, the binding and kinase response cannot be simultaneously captured by equilibrium Ising-type models (for various spatial structures), where the receptor activities $s_i$ are variable across the receptor complex (see SI Appendix).

\subsection{A nonequilibrium allosteric model captures both ligand binding and kinase activity for Tar}

\begin{figure*}[t]
    \includegraphics[width=.7\linewidth]{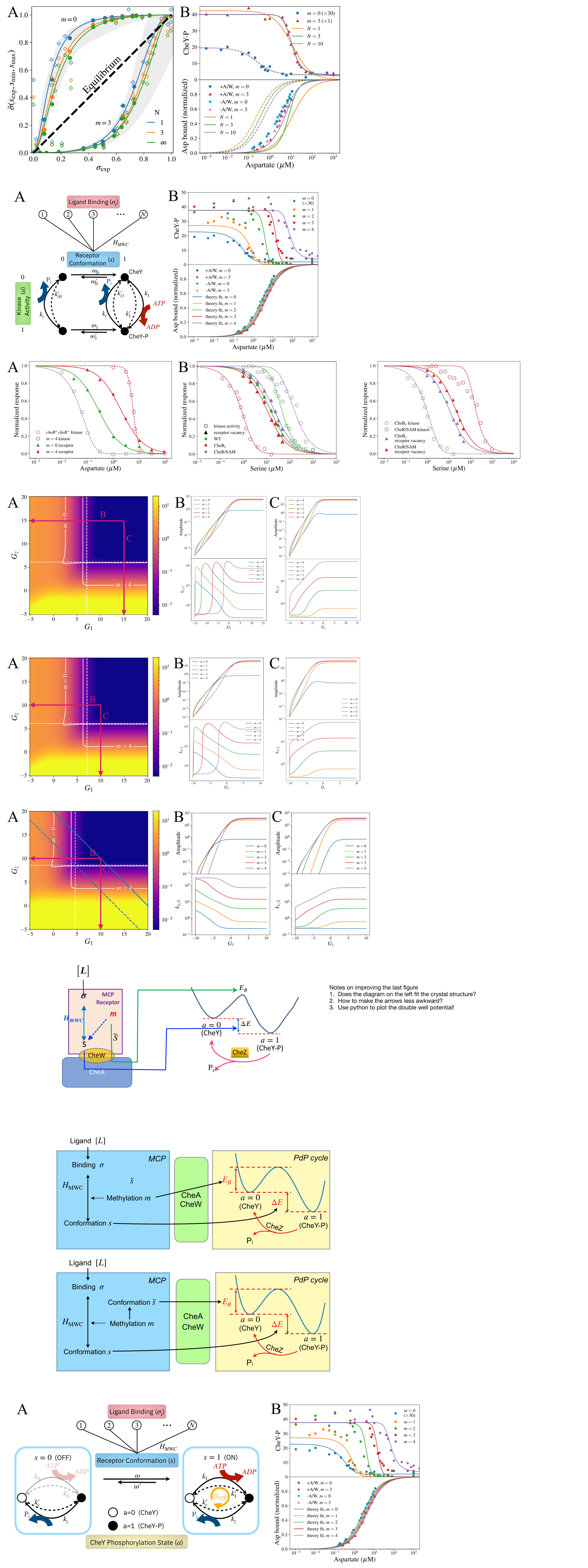}
    \caption{\label{4stateModel}
        The nonequilibrium allosteric model for the chemoreceptor.
        (A) Schematics of the model.
        The receptor conformation ($s$) is coupled to the ligand occupancy at $N$ binding sites ($\sigma_i$, $i=1,2,\dots, N$) via an equilibrium interaction described by the MWC model with the energy function $H_\mathrm{MWC}$.
        The receptor switches between ON and OFF states at rates $\omega$ and $\omega'$, which depend on the ligand occupancy.
        The CheY phosphorylation state $a$ is controlled by the phosphorylation-dephosphorylation (PdP) cycle driven by ATP hydrolysis.
        The phosphorylation is catalyzed by the active receptor ($s=1$) with a rate $k_1$ (we assume the phosphorylation rate to be negligible in the OFF state, $k_0=k_0'=0$).
        The dephosphorylation of CheY-P is catalyzed by CheZ with a rate $k_z$, independent of the receptor state.
        (B) The model captures both the measured CheY-P level (upper) and aspartate binding (lower) for Tar receptors. $N=10$, $G_\mathrm{ATP}=20$, $E_s=5$. All experimental data are from Amin and Hazelbauer~\cite{Amin2010Chemoreceptors}.
    }
\end{figure*}

The fact that equilibrium models cannot capture the binding and kinase response measurements indicates that the \emph{E. coli} chemotaxis signaling pathway operates out of equilibrium.
We now present a nonequilibrium allosteric model that extends the MWC model and can simultaneously capture the measured binding and response curves as well as their asymmetric shifts due to methylation.

In this model (pictured in Fig.~\ref{4stateModel}A), the receptor conformation is again represented as a binary variable $s$ with inactive (0) and active (1) states. It is coupled to ligand occupancy $\sigma_i$ by an equilibrium MWC model. The phosphorylation state of the response regulator CheY is represented by a new binary variable $a$, with dephosphorylated (0) and phosphorylated (1) states. The measured CheY-P concentration is proportional to the average phosphorylation variable $\expval{a}$.
The transitions between states $a=0,1$ involve phosphorylation-dephosphorylation (PdP) cycles (blue boxes), whose reaction rates depend on both the conformation $s$ and the methylation $m$.
The system operates out of equilibrium sustained by continuous ATP hydrolysis in the PdP cycle, and the average receptor behavior is determined by the probability distribution $P(a,s)$ of the nonequilibrium steady state.

When the receptor is active ($s=1$), it catalyzes the phosphorylation of CheY through the autophosphorylation of the histidine kinase CheA, which subsequently transfers the phosphate group to CheY. In our model, we do not consider all the detailed steps of phosphotransfer from ATP to CheY and assign an overall rate $k_1$ to describe the phosphorylation of CheY. For simplicity, we also assume that the phosphorylation rate is negligible when the receptor is inactive ($s=0$).
The dephosphorylation of CheY-P is catalyzed by CheZ, with a dephosphorylation rate $k_z$ independent of the receptor state. To ensure thermodynamic consistency, all reactions are reversible with reverse rates $k_1' = k_1 e^{-G_1}$ and $k_z'=k_ze^{-G_z}$.
The PdP cycle is driven by ATP hydrolysis with free energy $G_\mathrm{ATP}$ (measured in units of $k_BT$), which is dissipated per PdP cycle per CheY molecule. The rates obey the thermodynamic constraint:
\begin{equation}
    G_\mathrm{ATP} = \log \frac{k_1 k_z}{k_1' k_{z}'} = G_1 + G_{z},
    \label{G_ATP}
\end{equation}
where $G_1$ and $G_{z}$ are the free energy released by phosphorylation and dephosphorylation respectively. Note that $\frac{k_1 k_z}{k_1' k_{z}'}=e^{G_\mathrm{ATP}}\gg 1$, which clearly indicates a violation of detailed balance and the far-from-equilibrium nature of our model.

Motivated by structural details of the bacterial chemoreceptors (see Discussion), we assume that receptor methylation can affect kinase activity by changing the energy barrier in the phosphorylation reaction, which scales $k_1$ and $k_1'$ by the same factor. On the other hand, the dephosphorylation rate $k_z$ remains independent of $m$.
The ratio of the two rates can be written as: $k_1/k_z = \exp\qty[E_p - \Delta E_B(m)]$, where $E_p$ is a constant and $\Delta E_B(m)$ captures the allosteric effect of receptor methylation on the energy barrier for phosphorylation. Here, we adopt the simplest model where the energy barrier has a linear dependence on the methylation level, i.e. $\Delta E_B(m) = -\epsilon_m N m$. The energy difference $G_1=\log(k_1/k_1')$ is not affected by methylation.

The receptor conformation $s$ can switch between ON and OFF states at rates $\omega$ and $\omega'$, whose ratio is defined as $\alpha=\omega/\omega'$.
Throughout our analysis, switching is assumed to be much faster than the PdP reactions ($\omega,\omega' \gg k_1,k_z$), which means $s$ is in fast equilibrium with $\sigma$. Since the coupling between $s$ and $\sigma$ does not involve any nonequilibrium driving, we can describe it with the MWC free energy (Eq.~\ref{MWCenergy}). Matching the switching dynamics to the MWC model gives the rate ratio:
\begin{equation}\label{activityRatio}
    \alpha \equiv \frac{\omega}{\omega'} = e^{-E_s} \qty(\frac{1+[L]/K_i}{1+e^{-E_0} [L]/K_i})^{-N}.
\end{equation}

The measured CheY-P concentration is proportional to the steady-state average $\expval{a} = (1+ \mathcal{P})^{-1}$, where $\mathcal{P}$ is the probability ratio of being not phosphorylated versus phosphorylated:
\begin{equation}\label{4stateProbabilityRatio}
    \mathcal{P}   \equiv \frac{P(a=0)}{P(a=1)} = \frac{k_z + (k_z+k_1')\alpha}{k_{z}'+(k_{z}'+k_1) \alpha}.
\end{equation}
In the large dissipation (strong nonequilibrium driving) limit, the reverse reaction rates are negligible, i.e. $k_i'/k_i \to 0$.
With an additional assumption $e^{E_s} \gg 1$, the activity reduces to,
\begin{equation}\label{effectiveMWCactivity}
    \expval{a} \approx \left( 1+ e^{E_s-E_p - \epsilon_m N m} \qty(\frac{1+ [L]/K_i}{1+e^{-E_0} [L]/K_i} )^N \right)^{-1}.
\end{equation}
This expression has the same form as the MWC activity Eq.~(\ref{avActivity}) with a new effective energy for the active state $E_s^\text{eff} = E_s -E_p -\epsilon N m$, which depends on the methylation level $m$.
Given that the MWC model [Eq.~(\ref{avActivity})] has been successfully employed to fit the CheY-P curves (Fig.~\ref{equilibirum_test}B), we can expect a similar, if not better, fit from the nonequilibrium model.
Indeed, Fig.~\ref{4stateModel}B shows that the model successfully captures the CheY-P level of vesicle-bound Tar receptors at all five methylation levels~\cite{Amin2010Chemoreceptors}.

Since there is no feedback from the CheY phosphorylation state to the receptor, the average receptor conformation $\expval{s}$ and ligand occupancy $\expval{\sigma}$ are given by the MWC model, Eqs.~(\ref{avActivity}) and (\ref{avBinding}) respectively. For large $E_s$, we again have $\langle s \rangle \ll 1$ so that the binding curve closely resembles a Hill function with Hill coefficient $n_H=1$. Fig.~\ref{4stateModel}B shows that the same set of parameters used to fit kinase response also produces the correct binding curves, which is a significant improvement from the MWC model.

Why is the nonequilibrium model able to capture the asymmetric shift of binding and kinase response curves due to methylation while equilibrium models cannot?
In equilibrium models, methylation can only affect receptor behavior through thermodynamic control, i.e. by changing the energy difference between different states through $E_s$ and $E_0$. Since equilibrium interactions are always reciprocal (symmetric), the fact that ligand binding controls receptor activity means that there has to be an equally strong feedback from the receptor state to the ligand occupancy. Therefore,  this type of control shifts the binding and kinase response curves by similar amounts, inconsistent with experimental observations.
In the nonequilibrium model, however, the interaction between ligand binding and kinase activity can be non-reciprocal (asymmetric): the receptor complex acts as an enzyme that exerts kinetic control by changing the phosphorylation energy barrier $\Delta E_B(m)$. In the absence of strong feedback from the substrate (CheY) to the receptor complex, these changes in the energy barrier enable amplified shifts in the kinase response while maintaining modest shifts in binding response.
For this mechanism to function, however, the system must be driven out of equilibrium (in this case by continuous ATP hydrolysis). Energy dissipation is required to enable kinetic control, which has no impact on the steady-state distribution if the system is in equilibrium, and is necessary to maintain the asymmetric coupling between the phosphorylation state $a$ and the receptor state $s$.
As shown later, the response amplitude vanishes in the absence of energy dissipation.

\subsection{Further confirmation of the nonequilibrium \\model: Receptor conformational changes \\ and the Tsr receptor}
\begin{figure}[t]
    \includegraphics[width=\linewidth]{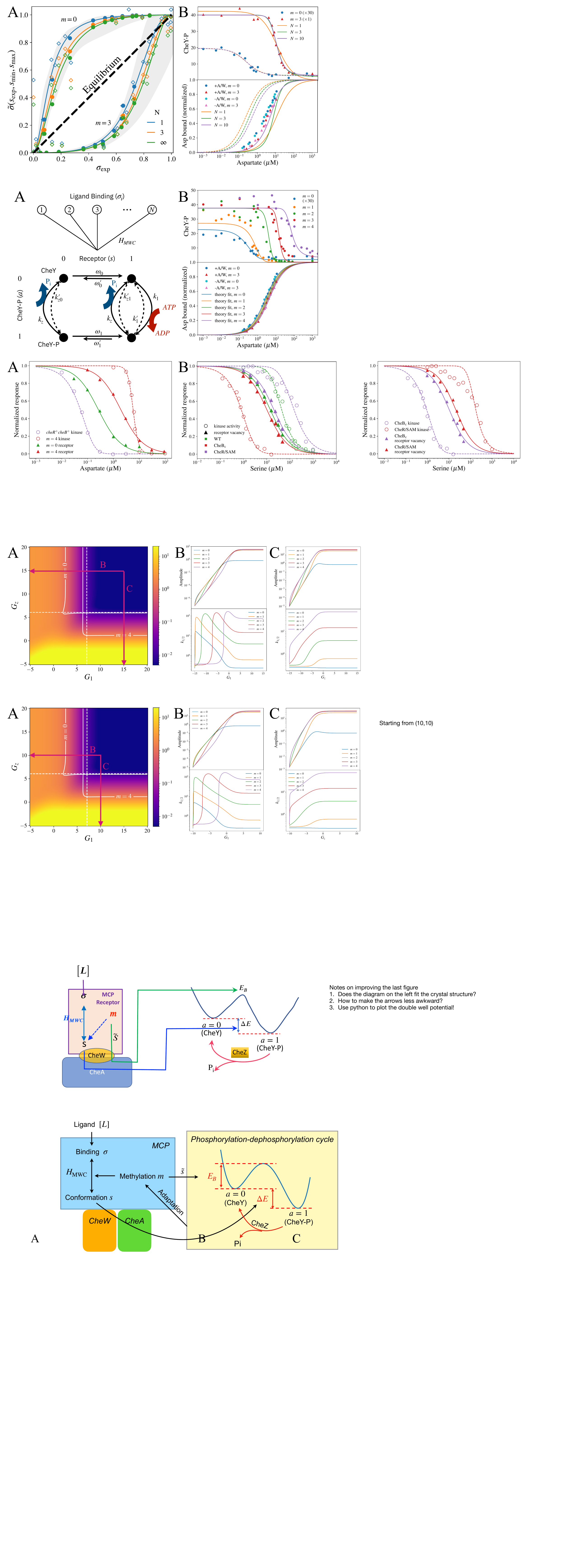}
    \caption{\label{model_fit}
        Additional experimental evidence in support of the nonequilibrium allosteric model.
        (A) Fits of the nonequilibrium model to \emph{in vivo} measurements of receptor conformational changes and CheY-P response by Vaknin and Berg~\cite{Vaknin07}. Conformation measurements were performed without CheA/CheW ($N \approx 1$).
        The $cheB^+ \, cheR^+$ measurement has an effective methylation level $m\approx 1$ \cite{shimizu2010modular}.
        $\blacktriangle$ indicates receptor conformation; $\circ$ represents kinase response.
        (B) Fits of the nonequilibrium model to \emph{in vitro} measurements of serine binding and CheY-P response of Tsr receptors by Levit and Stock~\cite{Levit02}. Measurements were done for WT cells in the absence of CheR or CheB (green) or  in the presence of $\text{CheB}_c$ (low methylation, red) or CheR and S-adenosylmethionine (high methylation, purple). $\blacktriangle$ indicates receptor vacancy $1-\sigma$; $\circ$ represents kinase response; the solid and dashed lines are fits with $N=10$.}
\end{figure}

In addition to explaining the asymmetric shifts in binding ($\sigma$) and kinase response ($a$) curves, our model also predicts different dose-response curves for the receptor conformation ($s$) and kinase response ($a$).
Vaknin and Berg measured Tar receptor conformation \emph{in vivo} by fusing fluorescent proteins to the C-termini of chemoreceptors and measuring the intensity of fluorescence resonance energy transfer (FRET) between receptor homodimers~\cite{Vaknin07}.
They measured the receptor conformation for mutants with fixed methylation states EEEE ($m=0$) and QQQQ ($m=4$) and the corresponding kinase response for QQQQ ($m=4$) and $cheR^+ cheB^+$ mutants.
The $cheR^+ cheB^+$ cells do not have a fixed methylation state, but their kinase response is known to be similar to QEEE mutants ($m=1$)~\cite{shimizu2010modular}.
The shift of the kinase response curve from $m=1$ to $m=4$ was found to be much more significant than the shift of the receptor conformation curve from $m=0$ to $m=4$ (Fig.~\ref{model_fit}A, circles versus triangles).
Note that the conformation measurements were carried out in the absence of kinase CheA and linker protein CheW, which leads to noncooperative responses ($N=1$).
Similar to the \textit{in vitro} case discussed above, the discrepancy between the dose-response curves suggests the necessity for a nonequilibrium allosteric model.
Indeed, as shown in Fig.~\ref{model_fit}A, the model simultaneously fits both the receptor conformation and kinase response curves for all mutants (solid and dashed lines).

Another abundant chemoreceptor in \textit{E. coli} is the serine receptor Tsr, which regulates CheY phosphorylation using the same microscopic mechanisms as the Tar receptor.
Levit and Stock~\cite{Levit02} measured the binding and kinase response for Tsr receptor complexes with three different receptor methylation levels (Fig.~\ref{model_fit}B).
This was achieved by expressing the WT receptor without CheR or CheB to fix its methylation level and subsequently increasing the methylation level by adding CheR and S-adenosylmethionine (SAM) or decreasing the methylation level by adding CheB\textsubscript{c}, the catalytic domain of CheB.
As the methylation level varied between these three conditions, the serine concentration required to inhibit kinase activity varied by more than two orders of magnitude, while the binding affinity only changed by two-fold (from $K_d=10\mu$M to $K_d=20\mu$M).
The asymmetric shift in binding and kinase response curves of Tsr receptors is similar to that found for Tar receptors, which is inconsistent with equilibrium models.
Once again, the discrepancy between these shifts can be fully captured by the nonequilibrium allosteric model (Fig.~\ref{model_fit}B, solid and dashed lines).

\subsection{The minimum dissipation and the critical effects of nonequilibrium driving}
\begin{figure*}[t]
    \includegraphics[width=0.75\linewidth]{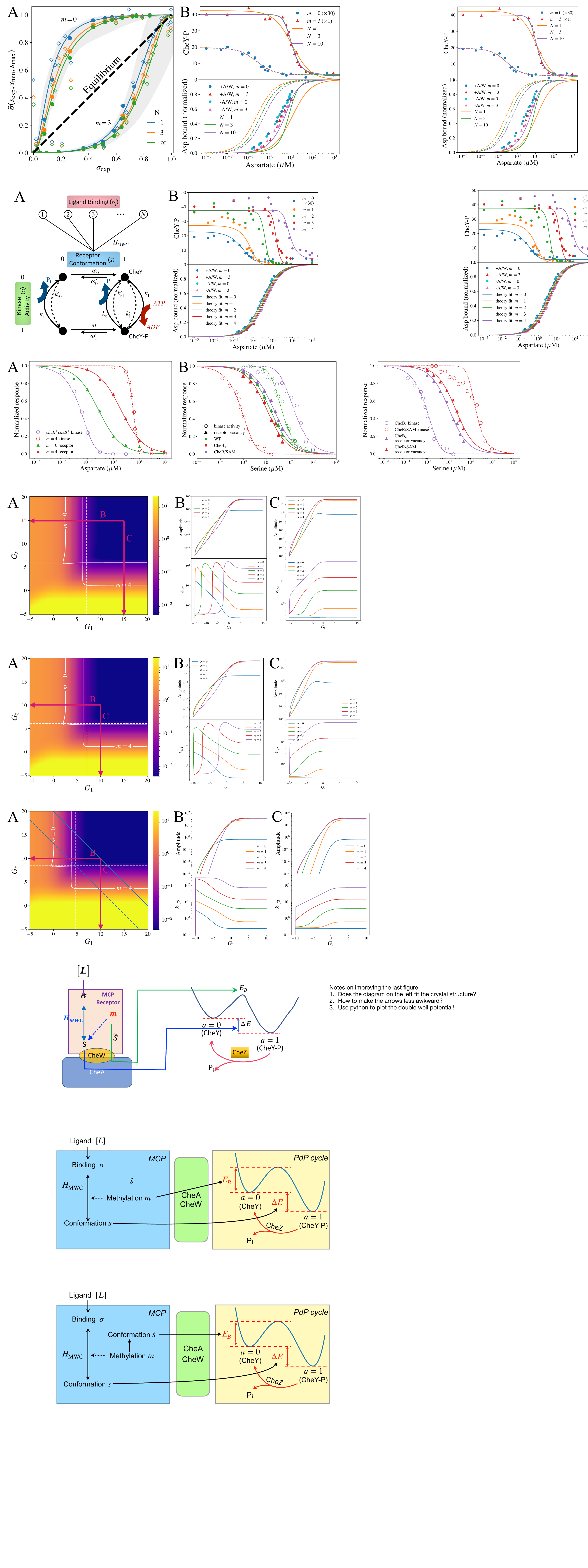}
    \caption{
    The dependence of the kinase response on the energy dissipation.
    (A) The difference $d(G_1, G_z)$ [Eq.~(\ref{activityDifference})] between the CheY phosphorylation level predicted by the nonequilibrium allosteric model at finite values of dissipation energies ($G_1$ and $G_z$) and that in the infinite dissipation limit.
    The solid white lines are contours at $d=d_\mathrm{th}=10^{-2}$ for $m=0$ and $m=4$. The white dashed lines are theoretical predictions of the operational thresholds $G_{1,\mathrm{th}}= -\log d_\mathrm{th}$ and $G_{z,\mathrm{th}}=-E_p+E_s-\log d_\mathrm{th}$.
    The solid and dashed blue lines indicate the physiological value $G_\mathrm{ATP}=20$ and the operational threshold $G_\mathrm{ATP}=13.2$, respectively.
    The purple arrows indicate paths taken in (B--C) toward the equilibrium limit.
    (B--C) The amplitude (upper) and the half-maximum aspartate level $k_{1/2}$ (lower) as the dissipation $(G_1, G_z)$ varies along the paths outlined in (A).
    }
    \label{Fig:dissipation_scaling}
\end{figure*}

Our results above show that strong nonequilibrium driving, which is provided by the free energy released during ATP hydrolysis ($G_\mathrm{ATP}$), is necessary to produce the experimentally observed large shifts in kinase response curves with changes in receptor methylation level (or other forms of modification to the receptor) while leaving ligand binding relatively unchanged.
This raises the question of how much energy dissipation is required for the system to exhibit behaviors that match the experimentally observed ligand binding and kinase response.

To address this question, we evaluate the difference between the average phosphorylation level $\expval{a}$ (proportional to the CheY-P concentration) and its value in the infinite dissipation limit $\expval{a}_\infty$ [given by Eq.~(\ref{effectiveMWCactivity})]  by expanding the full solution to the nonequilibrium model [Eq.~(\ref{4stateProbabilityRatio})] in terms of the reverse reaction rates $k_1'$ and $k_z'$. The leading order correction $\delta \langle a \rangle \equiv \expval{a}-\expval{a}_\infty$ is
\begin{equation}\label{activityExpansion}
    \begin{aligned}
        \delta \expval{a} = \frac{k_1 k_z \alpha (1+\alpha)}{(k_z + (k_1 + k_z) \alpha)^2} \qty(\frac{1}{P_\text{ON}}\frac{k_{z}'}{k_1}- P_\text{ON}\frac{k_1'}{k_z}),
    \end{aligned}
\end{equation}
where $P_\text{ON} = \alpha/(1+\alpha)$ is the probability of the receptor being in the ON state ($s=1$).
The two terms in the parenthesis have clear physical meanings: they quantify the irreversibility of the PdP cycle by comparing the reaction rates along or against the nonequilibrium driving (yellow arrow in Fig.~\ref{4stateModel}A).
The first term $k_z'/k_1$ is the ratio between the reverse dephosphorylation rate $k_z'$ (counterclockwise transition) and the phosphorylation rate (clockwise transition) originating from the CheY state ($a=0$). Similarly, the second term $k_1'/k_z$ is the ratio between reverse phosphorylation (counterclockwise) and dephosphorylation (clockwise) reactions from the CheY-P state ($a=1$). Since phosphorylation only occurs in the ON state, while dephosphorylation occurs independent of the receptor state, each term is weighted by an appropriate factor of $P_\text{ON}$. To make each of these terms small requires large $G_z$ and $G_1$ (and hence large $G_\mathrm{ATP}$) to drive the PdP cycle sufficiently irreversibly so that the system behaves close to the infinite-dissipation limit.

To quantify the minimum required dissipation $G_\mathrm{ATP}$, we start with Eq.~(\ref{activityExpansion}) and focus on the extreme cases $m=0$ and $m=4$, whose response curves envelop those of intermediate $m$.
For $m=0$, the CheY-P response is suppressed: the phosphorylation rate $k_1 = k_z \exp(E_p)$ is much less than the dephosphorylation rate $k_z$ ($E_p \approx -4$ for our best fit in Fig.~\ref{4stateModel}), so the first term in Eq.~(\ref{activityExpansion}) dominates. We want the error to be small compared to the maximal infinite-dissipation activity, i.e.,  $|\delta \expval{a}/\max (\expval{a}_\infty)| < d_{th}$ for some loss threshold $d_{th}$, where $\max(\expval{a}_\infty) = (1+e^{Es} k_z/k_1)^{-1}$. Since $E_s$ is large and therefore $\alpha$ is small, this leads to $k_{z}'/(k_1 \alpha) < d_{th}$, or
\begin{equation}\label{gz1Threshold}
    G_{z} > (-E_p+E_s) - \log(d_{th}).
\end{equation}
Conversely, for $m=4$, the CheY-P response is amplified by the kinetic barrier shift due to methylation, and we have $k_1 = k_z \exp(E_p + 4 \epsilon_m N) \gg k_z$. Therefore, the second term in Eq.~(\ref{activityExpansion}) dominates. Again requiring this term is small compared to the maximal infinite-dissipation activity leads to $k_{1}'/k_1 < d_{th}$, or
\begin{equation}\label{g1Threshold}
    G_1 > - \log(d_{th}).
\end{equation}
Adding these two inequalities gives an operation threshold for $G_\mathrm{ATP}$,
\begin{equation}\label{gAtpThreshold}
    G_\mathrm{ATP} >G_{\min}= -E_p + E_s - 2 \log(d_{th}).
\end{equation}
From our fits, $E_p-E_s \approx -4$, so using a threshold of $d_{th}=10^{-2}$ leads to $G_{\min} = 13.2$, which is the minimum dissipation energy needed to drive the system to have the experimentally observed behaviors shown in Fig.~\ref{4stateModel}B.

As given in Eq.~(\ref{G_ATP}), the total dissipation energy ($G_\mathrm{ATP}$) can be decomposed into two parts: $G_\mathrm{ATP}=G_1+G_z$ where $G_1$ and $G_z$ are used to suppress the reverse reactions for the phosphorylation and the dephosphorylation processes, respectively. In the infinite dissipation limit with $G_1, G_z\rightarrow \infty$, all the reverse reactions can be neglected, and we obtain the exact expression for the kinase activity given in Eq.~(\ref{effectiveMWCactivity}), which agrees quantitatively with the experiments. For finite values of $G_1$ and $G_z$, the behavior of the model can be different.
The difference between the phosphorylation levels at finite and infinite dissipation can be quantified by
\begin{equation}\label{activityDifference}
    d(G_1,G_z) = \sum_{m=0}^4 w_m \sqrt{\expval{ \qty[\expval{a}(m,[L]) -\expval{a}_\infty (m,[L])]^2
        }},
\end{equation}
where the weight $w_m = 1/\expval{a}_\infty(m,0)$ normalizes curves by the infinite-dissipation maximal activity at the corresponding methylation state, and the average is calculated by sampling uniformly over the activity range of $\expval{a}_\infty$.
Fig.~\ref{Fig:dissipation_scaling}A shows $d(G_1,G_z)$ for different values of $G_1$ and $G_z$.
As expected, the difference is small in the upper right region when both $G_1$ and $G_z$ are sufficiently large, which defines the operational regime of the chemoreceptor sensory system.
The estimated thresholds Eqs.~(\ref{gz1Threshold}) and (\ref{g1Threshold}), shown by white dashed lines, accurately predict when the models have quantitatively similar kinase activity. Moreover, they are indeed limited by methylation levels $m=0$ and $m=4$ as shown by the contours of the output discrepancy for individual methylation levels (white solid lines).
The blue dashed line shows the operational threshold $G_{\min} = 13.2$, where the difference is only small for a particular pair of $(G_1, G_z)$. In contrast, for a typical physiological value $G_\mathrm{ATP}=20$~\cite{tran_changes_1998} (blue solid line), $(G_1, G_z)$ is allowed to vary within a much larger range without deviating from the experimentally observed kinase response.

Next, we investigate the role of energy dissipation in the signaling pathway. In particular, we find that it enhances the amplitude and the sensitivity range of the response.
To illustrate this, we tune the nonequilibrium driving $G_\mathrm{ATP}$ and track how it affects the CheY-P response amplitude and the half-maximum aspartate concentration ($k_{1/2}$) for each methylation level.
To demonstrate their typical behaviors, we consider two paths shown by the purple arrows in Fig.~\ref{Fig:dissipation_scaling}A, with the corresponding responses shown in Fig.~\ref{Fig:dissipation_scaling}B and C.
Both paths start from $G_\mathrm{ATP} = 20$ ($G_1 = G_z = 10$) and move towards the equilibrium limit ($G_\mathrm{ATP}=0$) by decreasing $G_1$ with fixed $G_z$ and vice versa.
In each case, the amplitude decreases to zero as the dissipation approaches zero. Indeed, the receptor state affects CheY phosphorylation through kinetic control, which has no effect in the equilibrium limit.
Furthermore, non-zero dissipation enables the large spread in CheY-P response curves. Because our model has no feedback from the phosphorylation state to the receptor, and therefore no response in equilibrium, the response curves snap together suddenly when $G_\text{ATP}=0$. For the response to be nonzero in the equilibrium limit, there has to be feedback from CheY to the receptor. Thus, the relation Eq.~(\ref{activityLinearRelation}) holds and the spread between kinase response curves must be comparable to that for binding curves.
Interestingly, the $k_{1/2}$ scaling is qualitatively different depending on the path taken to equilibrium: varying $G_1$ leads to both non-monotonic behavior, with $k_{1/2}$ initially increasing for each $m$ before decreasing dramatically near equilibrium. Conversely, varying $G_z$ produces monotonic scaling with an initial gradual decrease in the sensitivity range before a similar collapse at equilibrium.

The dissipation energies $G_1$ and $G_z$ are connected to biochemical parameters via the following relationships,
\begin{equation}
    G_1 = G^0_\mathrm{ATP} - G^{0}_\mathrm{CheY-P} + \log \frac{[\mathrm{ATP}]/[\mathrm{ATP}]_0}{[\mathrm{ADP}]/[\mathrm{ADP}]_0}
\end{equation}
and
\begin{equation}
    G_z = G^{0}_\mathrm{CheY-P} + \log [\mathrm{P_i}]/[\mathrm{P_i}]_0,
\end{equation}
where $G^0_\mathrm{ATP}$ and $G^{0}_\mathrm{CheY-P}$ are the phosphate bond energies for ATP and CheY-P, respectively and $[\mathrm{ATP}]_0$, $[\mathrm{ADP}]_0$ and $[\mathrm{P_i}]_0$ are the reference concentrations.

A possible scheme to test our model could be to measure kinase dose-response curves while tuning the [ATP]/[ADP] ratio to change $G_1$. For a fixed value of $G_z$ within the operational regime, the response amplitude and sensitivity range will decrease as the [ATP]/[ADP] ratio decreases. For example, if $G_z=10$, as is the case in Fig.~\ref{Fig:dissipation_scaling}B, the response amplitude for $m=4$ will decrease by about 5\% when the [ATP]/[ADP] ratio decreases by three orders of magnitude~\footnote{Changes in ATP concentration of nearly this magnitude (400-fold) have been used in studies of the processive motion of molecular motors \cite{gebhardt2006myosin}}.
Note that this estimate is nearly the worst-case scenario, since our choice of initial $(G_1,G_z)$ has $G_z$ close to its threshold value. If the true $G_z$ is larger than 10, a much larger reduction in amplitude will occur for the same decrease in the [ATP]/[ADP] ratio (e.g., 51\% for $G_z = 13$). Furthermore, since our model is coarse-grained (for example, it ignores intermediate steps in the phosphotransfer between CheA and CheY), the predicted dissipation rate is smaller than that of the real system \cite{yu2021inverse,yu_state-space_2022}.
Therefore, the real kinase response may be more susceptible to changes in ATP concentration than the model predicts. Another possible scheme to test our model is to vary $G_z$ by controlling the inorganic phosphate concentration $[P_i]$ as shown in Fig.~\ref{Fig:dissipation_scaling}C.
Quantitative measurements of the amplitude and sensitivity range using one of the proposed schemes above may help more precisely pinpoint the operational regime of chemoreceptor signaling complexes.

\section{Summary and Discussion}
In this paper, we have developed a nonequilibrium allosteric model of bacterial chemoreceptors motivated by the asymmetric shifts in ligand binding and kinase response curves caused by receptor methylation, which has been a long-standing puzzle in the field.
The model explains all existing measurements of ligand binding, receptor conformation, and kinase response within a unified framework that takes into account both allosteric interactions within the receptor-kinase complex as well as the nonequilibrium phosphorylation-dephosphorylation reaction kinetics driven by energy dissipation.

How intracellular energy is used to drive information and signaling processes in living cells is a fundamental question in biological physics. Recently, much progress have been made in elucidating the critical effects of energy dissipation in a wide range of cellular functions such as the ultrasensitive bacterial flagellar motor switch~\cite{tu2008nonequilibrium}, accurate sensory adaptation~\cite{lan_energyspeedaccuracy_2012}, error correction~\cite{murugan_speed_2012, sartori_thermodynamics_2015}, gene expression control~\cite{estrada_information_2016}, and biochemical oscillation and synchronization~\cite{cao_free-energy_2015,zhang_energy_2020}. Here, by combining experimental data and theoretical modeling,  we show that strong dissipation, fueled by ATP hydrolysis, is necessary for enhancing the response amplitude and sensitivity range of the sensor-kinase signaling process.

Below, we discuss the possible microscopic mechanism underlying the nonequilibrium allosteric model, the generality of our model, and some future directions to extend our model.

\subsection{Possible microscopic mechanism: ligand binding and methylation cause different conformation changes}

\begin{figure}[t]
    \includegraphics[width=\linewidth]{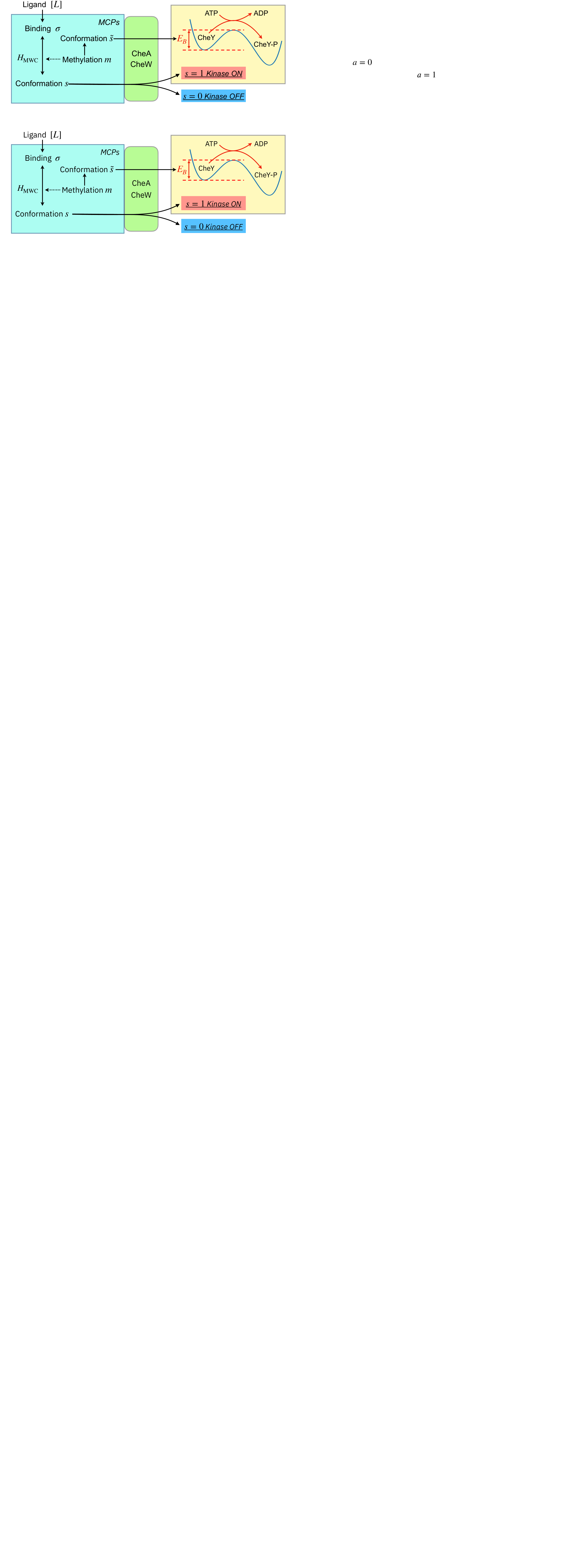}
    \caption{Illustrative summary of the nonequilibrium allosteric model. The receptor can undergo two separate conformational changes $s$ and $\tilde s$, in response to ligand binding and methylation, respectively. Each conformation affects the phosphorylation of CheY through CheA, but in different ways:
        $s$ controls switching between the kinase OFF state, with no phosphorylation of CheY, and ON state, which has a finite energy barrier for phosphorylation. In the ON state, $\tilde s(m)$ lowers the energy barrier $E_B$ (by different amounts $\Delta E_B(m)$ depending on the methylation level) to increase the phosphorylation rate. The phosphorylation reaction, controlled through these two mechanisms, together with dephosphorylation catalyzed by CheZ (not shown) determine the steady state phosphorylation level of CheY.
    }
    \label{cartoon}
\end{figure}

The proposed microscopic mechanism underlying the nonequilibrium allosteric model is summarized in Fig.~\ref{cartoon}.
The average phosphorylation state of the response regulator CheY is controlled by the methyl-accepting chemotaxis proteins (MCP, blue box) through CheA and CheW (green box). The MCP can undergo multiple distinct conformational changes (denoted by $s$ and $\tilde s$), each of which affects the phosphorylation rate of CheY in different ways.

The conformation $s$ is predominantly controlled by ligand binding ($\sigma$) via an equilibrium mechanism captured by the MWC energy function $H_\mathrm{MWC}$ (methylation may have a minor effect on $s$ through $K_i$, $E_s$, or both).
When $s=0$, the receptor is in the OFF state with almost no kinase activity, which is equivalent to having a very large energy barrier ($E_B\gg 1$).
When $s=1$, the receptor is in the ON state with a finite (but not unique) kinase activity, whose intensity depends on the methylation $m$.
Given that the receptor methylation sites are away from the kinase CheA, we hypothesize that a different conformation state $\tilde{s}$ mediates the allosteric control of methylation on the kinase activity.
In particular, increasing $m$ induces a conformational change $\tilde{s}(m)$ that lowers the energy barrier $E_B$ and thereby increases the phosphorylation level of the ON state.
The change in the barrier height is exactly $\Delta E_B(m)$ as defined in our nonequilibrium allosteric model.
It is important to note that while $s$ acts like a binary switch that can be described by a two-state model, $\tilde{s}$ may represent multiple or even a continuous spectrum of conformations, which lead to different kinetic rates or equivalently different barrier heights in the CheY phosphorylation reaction.

The idea that ligand binding and receptor methylation induce different conformational changes in the chemoreceptor has been suggested in the experimental literature~\cite{le_moual_conformational_1998,murphy_hydrogen_2001} and is supported by recent \emph{in vivo} crosslinking measurements of Tsr receptors~\cite{flack_structural_2022}, which found differing receptor structure changes in response to ligand binding and methylation. In particular, the change in conformation (quantified by the fraction of crosslinking products) induced by ligand binding is about twice as large as the corresponding change between methylation levels $m=0$ and $m=4$.

Another line of evidence comes from the dynamical properties of the cytoplasmic helical domains of nanodisc-inserted receptors, which have been measured using electron paramagnetic resonance (EPR) spectroscopy~\cite{bartelli_differential_2015,bartelli_bacterial_2016}.
It was found that increasing methylation reduces the mobility of the helical structure but preserves the mobility difference between companion helices and among different functional regions. In contrast, the conformation change induced by ligand binding has no detectable effect on helical mobility.
Since helical mobility is primarily affected by methylation, similar to the conformation $\tilde{s}$ in the model, we conjecture that helical dynamics may play a role in tuning the phosphorylation energy barrier in the receptor ON state.

These findings directly support the idea of distinct conformational changes ($s$ and $\tilde{s}$) used in our model, which leads to the following emerging picture for chemoreceptors operation: ligand binding induces a large conformational change that switches the receptor from active to inactive; methylation leads to more subtle conformational and dynamical changes that modulate the kinetic properties of the active state.
More structural insight into the conformation states is needed to more concretely disentangle the differing effects of ligand binding and methylation. For example, repeating recent FRET measurements of nanodisc-inserted Tar receptors~\cite{Gordon2021Concerted} at different methylation levels would help to clarify this picture.

\subsection{Generality of the nonequilibrium allosteric model: application to other two-component systems}
\begin{figure}[t]
    \includegraphics[width=\linewidth]{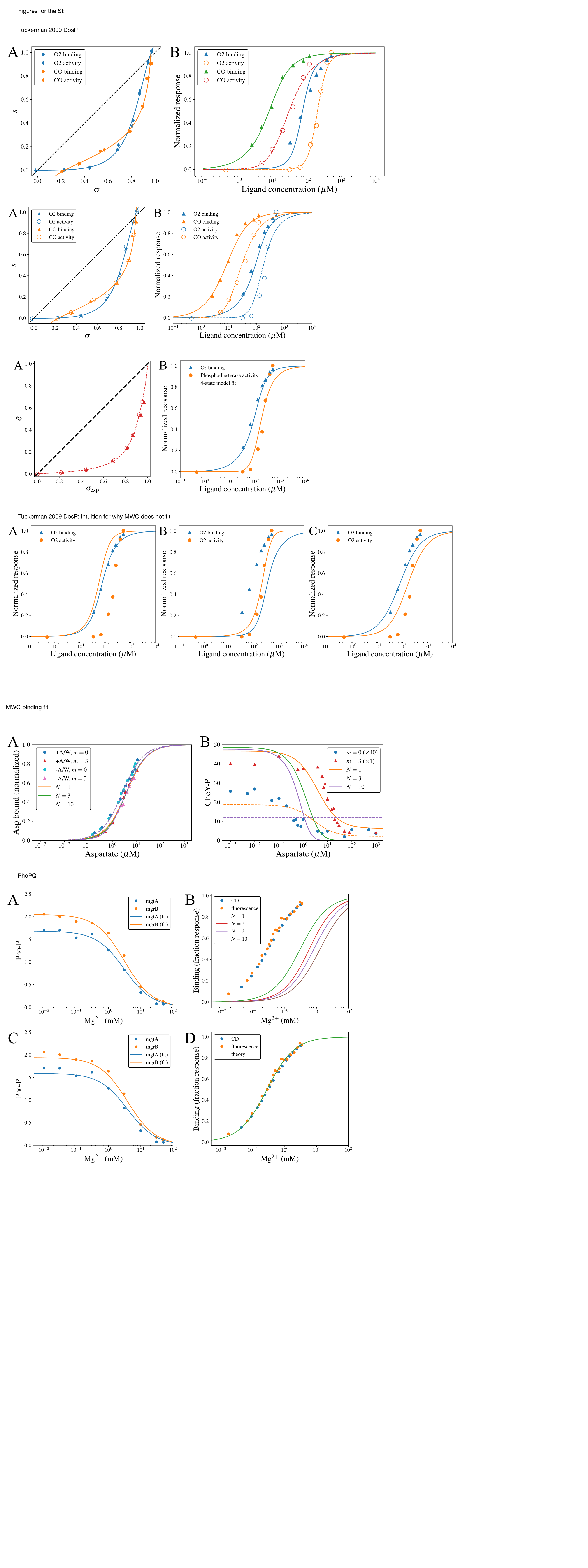}
    \caption{
        The nonequilibrium allosteric model captures the ligand binding and phosphodiesterase activity of \textit{E. coli} DosP system for O\textsubscript{2} sensing.
        (A) A parametric test plotting the inferred binding $\tilde\sigma$ (from the equilibrium MWC model) against measured $\sigma$.
        The triangles and circles are inferred $(\sigma, \tilde\sigma)$ pairs from binding and activity measurements, respectively. The dashed line is a guide for the eye obtained by fitting raw data to a Hill function.
        (B) The oxygen binding (blue) and DosP  phosphodiesterase activity (orange) and the best fit by our model (solid lines). $N=4$ is used since DosP forms tetramers~\cite{yoshimura_relationships_2003}.
        See SI Appendix for details of the fit.
    }
    \label{DosP}
\end{figure}

Beyond chemotaxis, our model should be broadly applicable to other two-component signaling systems, which have been studied extensively in \emph{E.~coli} and other bacteria~\cite{krell_bacterial_2010}.
Here, we illustrate these applications with the \textit{E.~coli} oxygen sensor protein DosP~\cite{delgado-nixon_dos_2000,sasakura_characterization_2002,tuckerman_oxygen-sensing_2009}.
Increasing oxygen concentration promotes DosP's ability to hydrolyze cyclic di-GMP, an important bacterial second messenger that triggers downstream responses~\cite{romling_cyclic_2013,jenal_cyclic_2017}.
Previous work measuring both oxygen binding and phosphodiesterase activity of DosP as functions of oxygen concentration reported that the two measurements cannot be consistently explained by an equilibrium model~\cite{tuckerman_oxygen-sensing_2009}.
Indeed, inconsistency with equilibrium MWC models can be more systematically demonstrated using the parametric test introduced in this work (Fig.~\ref{DosP}A).
A slightly generalized version of the nonequilibrium model (see SI Appendix) is able to capture both the binding and the phosphodiesterase activity curves. As shown in Fig.~\ref{DosP}B, it captures both the sharpness and sensitivity range difference between the binding and activity curves.
In contrast, as established by the parametric test, the MWC model can only capture either but not both response curves (see SI Appendix).

Since many of these two-component systems involve the hydrolysis of energy-rich molecules such as NTP, we expect simultaneous measurements of ligand binding and downstream activity dose-response curves to reveal potential inconsistencies with equilibrium models that are widely adopted.
Indeed, inconsistency with equilibrium models has also been reported for the FixL/FixJ system~\cite{sousa_memory_2007}.
Large discrepancies between the half-maximal concentrations for binding and activity have also been observed for the PhoP/PhoQ system~\cite{vescovi_characterization_1997}, suggesting that it operates out of equilibrium.
Overall, the theoretical framework presented here enables a deeper understanding of the mechanism of two-component systems by fitting simultaneously to measurements of binding and enzymatic activity.

More broadly, we expect that the nonequilibrium allosteric model developed here may be useful for understanding other biological signaling systems such as the G-protein coupled receptor (GPCR) signaling pathways if the receptor binding and the output activity can be measured and modeled simultaneously.

\subsection{Future directions: structure-based nonequilibrium models for large protein clusters}

Besides application to other two-component systems, another extension of this work is to better incorporate structural information.
In the introduction, we noted that the MWC model has two major simplifications: it is equilibrium and it neglects the rich spatial structure of large protein complexes. We have addressed the first of these shortcomings by developing a minimal nonequilibrium model of allostery that is sufficient to explain most of the existing measurements of chemotactic signaling.
Looking forward, incorporating spatial structure into the nonequilibrium model will allow us to probe one of the most important questions in biology: how does structure determine function? As the structural information for these complexes becomes available thanks to the recent development of high-resolution techniques such as cryo-electron tomography, the challenge is how this structural information can be used to understand the functions of these large complexes. \emph{E. coli} chemoreceptor clusters form a beautiful hexagonal lattice structure by tiling core units made up by two trimers of receptor dimers bound with a CheA dimer and two CheW~\cite{Briegel2102Bacterial,liu_molecular_2012}. Developing a nonequilibirum model containing these spatial details will be essential for understanding a variety of recent experimental discoveries including asymmetric activity switching times~\cite{keegstra_phenotypic_2017,keegstra_near-critical_2022}, emergent asymmetric coupling in mixed receptor complexes~\cite{yuhai2003coupling}, and slow logarithmic decay in activity on long timescales in saturating ligand~\cite{frank_prolonged_2013}.

\vspace{1em}

\begin{acknowledgments}
    This work is supported in part by National Institutes of Health grant R35GM131734 (to Y.~T.).
    Q.~Y.~acknowledges the IBM Exploratory Science Councils for a summer internship during which part of the work was done.
    G.~L.~H. acknowledges his NIH MERIT Award (R37GM29963).
\end{acknowledgments}

\bibliography{ref_chemotaxis}
\end{document}


\title{Supplemental Information:  Resolving the binding-kinase discrepancy in bacterial chemotaxis: A nonequilibrium allosteric model and the role of energy dissipation}

\author{David Hathcock}
\email{These two authors contributed equally}
\affiliation{IBM T.~J.~Watson Research Center, Yorktown Heights, NY 10598}

\author{Qiwei Yu}
\email{These two authors contributed equally}
\affiliation{IBM T.~J.~Watson Research Center, Yorktown Heights, NY 10598}
\affiliation{Lewis-Sigler Institute for Integrative Genomics, Princeton University, Princeton, NJ 08544}

\author{Bernardo A.~Mello}
\affiliation{International Center of Physics, Physics Institute, University of Brasilia, Brasilia, Brazil}

\author{Divya N.~Amin}
\affiliation{Department of Biochemistry, University of Missouri, Columbia, MO 65211}

\author{Gerald L.~Hazelbauer}
\affiliation{Department of Biochemistry, University of Missouri, Columbia, MO 65211}

\author{Yuhai Tu}
\affiliation{IBM T.~J.~Watson Research Center, Yorktown Heights, NY 10598}

\maketitle

\section{Testing equilibrium allosteric models}

\subsection{MWC model: fitting ligand binding leads to inconsistent kinase activity prediction}
\begin{figure}
    \centering
    \includegraphics[width=0.8\linewidth]{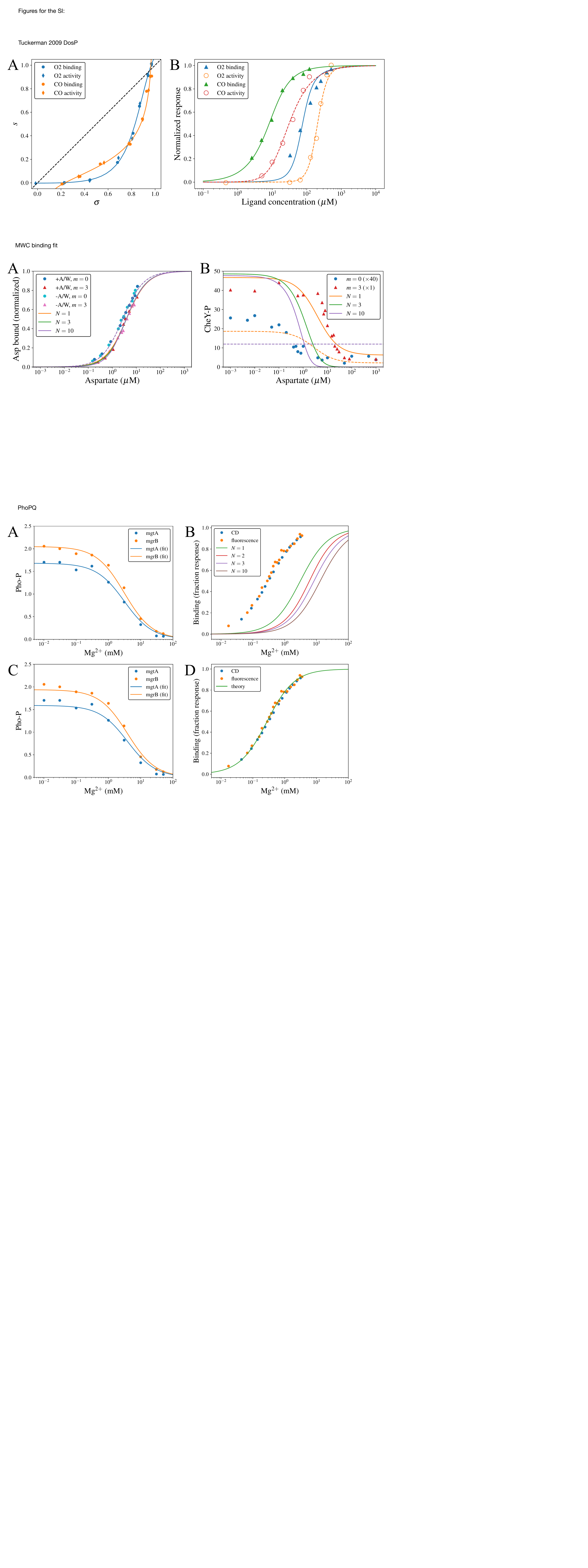}
    \caption{
        The measurements by Amin and Hazelbauer~\cite{Amin2010Chemoreceptors} are inconsistent with the MWC model. Here, we fit the binding curves and use the fitting parameters to predict kinase activity.
    }
    \label{fig:Amin binding}
\end{figure}

In similar spirit to Fig.~1B of the main text, we use the MWC model to fit the binding curves and plot the resulting kinase activity curves. The results are shown in Fig.~\ref{fig:Amin binding}, which is another demonstration that the MWC model is inconsistent with the data.

\subsection{Linear dependence between additional degrees of freedom}
In the main text [Eq.~(7)], we argued that coupling an additional equilibrium degree of freedom (e.g. kinase activity) to the MWC model produces a response that is linearly related to the MWC activity. Here we show this result is quite general and holds for any chain of binary variables coupled via the following Hamiltonian,
\begin{equation}
    H = (-\mu + E_{0,1}s_1) \sum_{i=1}^N \sigma_i + \sum_{i=1}^{M-1} (E_i s_i +E_{i,i+1} s_i s_{i+1}) + E_M s_M.
\end{equation}
As in the main text, $\sigma_i=0,1$ denotes the receptor occupancy, which depends on ligand concentration $[L]$ and dissociation constant $K_i$ through the chemical potential $\mu = \log([L]/K_i)$. The variables $s_i = 0,1$ are activities for various components of the system. For example, these might represent different conformational changes along the receptor body or kinase-controlling tip as well as conformational changes of the kinase itself, each coupled in a chain via equilibrium mechanisms. For each $s_i$, $E_i$ is the energy difference between the active and inactive states, and $E_{i,i+1}$ is the coupling energy between neighboring conformations.

We can relate the average states of neighboring conformational degrees of freedom using conditional probability:
\begin{equation}
    \begin{split}
        \langle s_{j+1} \rangle &= \langle s_{j+1}|s_{j}=0 \rangle P(s_j=0) + \langle s_{j+1}|s_j=1 \rangle P(s_j=1) \\
        & =  \langle s_{j+1}|s_{j}=0 \rangle (1- \langle s_j \rangle) + \langle s_{j+1}|s_j=1 \rangle \langle s_j \rangle\\
        & = C_0 + C_1 \langle s_j \rangle,
    \end{split}
\end{equation}
where $C_0$, $C_1$ are constants independent of $\mu$. The final step relies on the interactions between activities being equilibrium, so that the conditional expectation $\expval{s_{j+1}|s_j=0,1}$ is independent of $\mu$. For nonequilibrium models, including the model in the main text, this is not generally true: the conditional expectation can be a function of $\expval{s_j}$ and therefore maintain $\mu$-dependence. Given the linear relationship between $\expval{s_{j+1}}$ and $\expval{s_j}$ in equilibrium models, the response curves must be identical after normalization,
\begin{equation}
    \frac{\expval{s_{j+1}}_{\max} - \expval{s_{j+1}}}{\expval{s_{j+1}}_{\max}-\expval{s_{j+1}}_{\min}}=    \frac{\expval{s_{j+1}}_{\max} - \expval{s_{j+1}}}{\expval{s_{j+1}}_{\max}-\expval{s_{j+1}}_{\min}}.
\end{equation}
Since linear proportionality applies to any pair of neighboring activities, it extends to the entire chain. Therefore, if a system is governed by equilibrium interactions between neighboring conformational states, any measurement of activity can be captured by an effective MWC model.

\subsection{Ising models fail to explain both ligand binding and kinase response}
\begin{figure}
    \centering
    \includegraphics[width=0.35\linewidth]{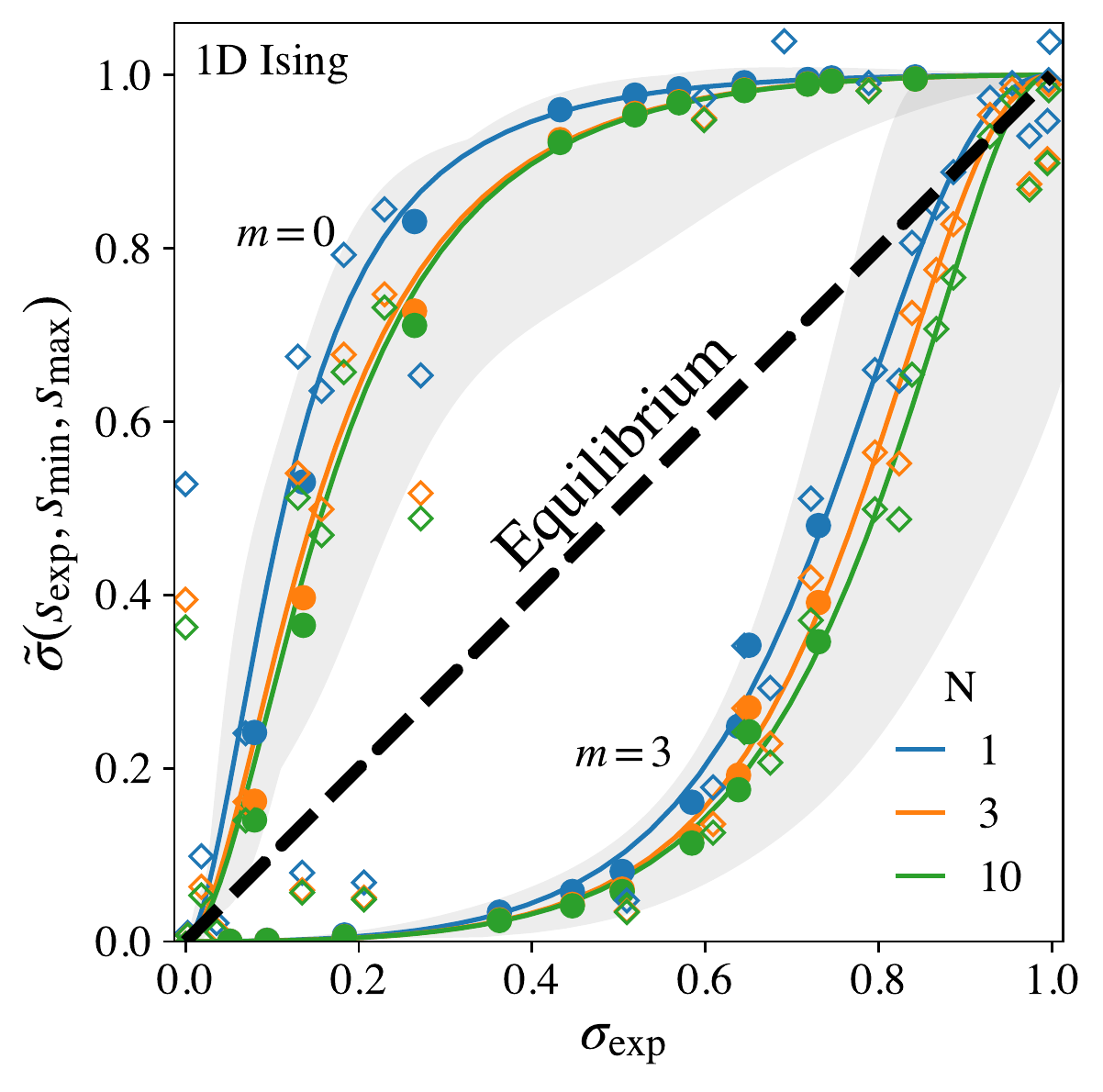} \,     \includegraphics[width=0.35\linewidth]{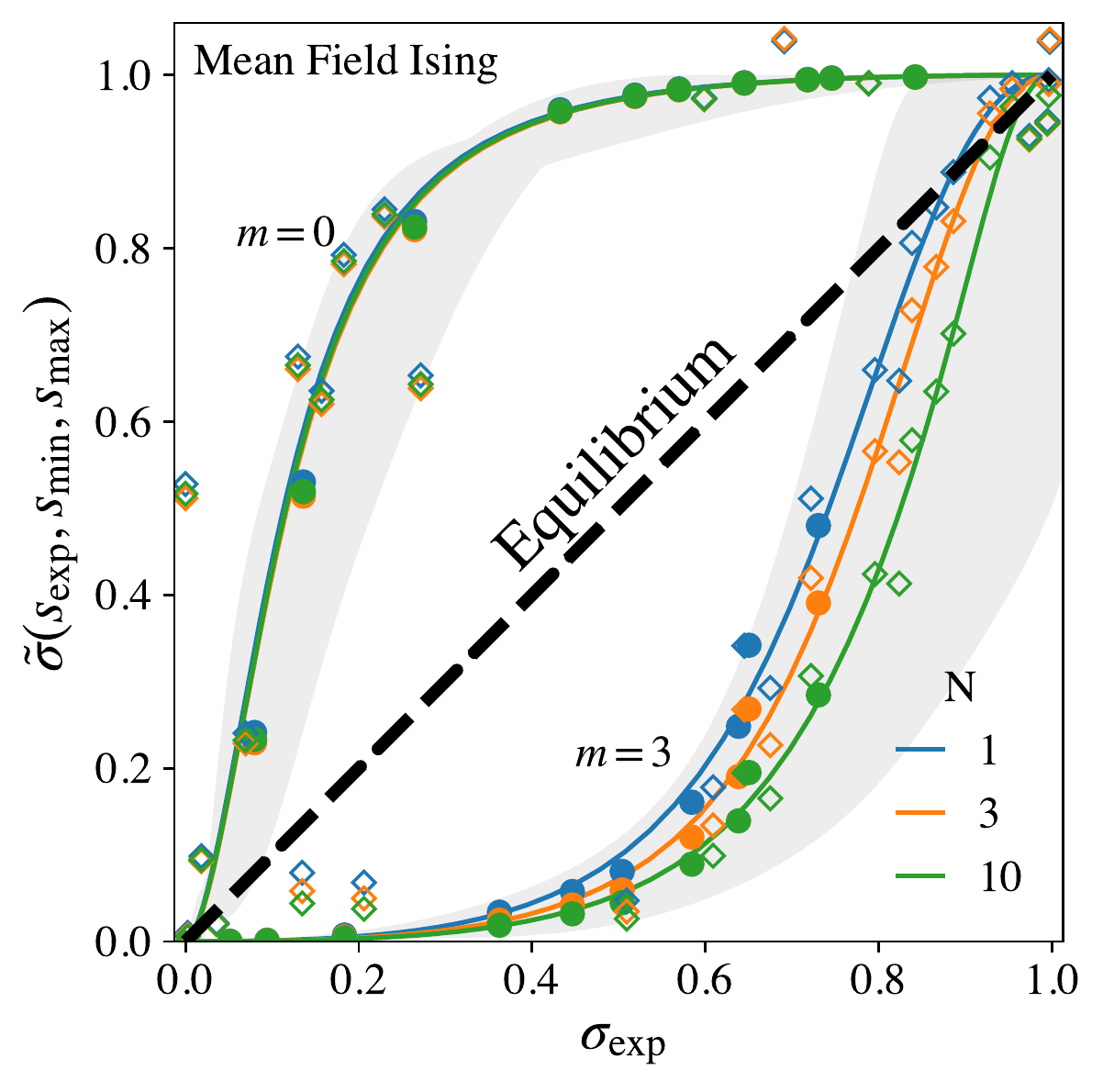}
    \caption{
        Measured aspartate binding and kinase activity of Tar-containing signaling complexes~\cite{Amin2010Chemoreceptors} are inconsistent with equilibrium Ising models.
        Parametric test plotting the occupancy inferred from kinase activity measurements $\tilde \sigma(s_{\rm{exp}}, s_{\max}, s_{\min}|J,N)$ against aspartate binding measurements $\sigma_\mathrm{exp}$ for (A) 1D Ising models and (B) Mean-field Ising models. In each panel, we show the test for $N=1$,$3$, and $10$. Aspartate binding measurements ($\newmoon$), CheY-P concentration measurements ($\Diamond$), and Hill function fits (solid lines) are transformed using coupling $J = 5$. The gray shaded area shows the envelope of $95\%$ confidence regions from Hill function fits for each $N$ and for $J$ in the range $[0,10]$. Curves are identical to the MWC parametric test for $N=1$ and qualitatively similar for $N>1$ (compare to Fig.~1, main text). The transformed data lie well off the diagonal, indicating the presence of nonequilibrium driving.
    }
    \label{fig:IsingTests}
\end{figure}

Ising models are also frequently used to study the behavior of chemoreceptor clusters~\cite{Mello03a, Lan2011Adapt}. In contrast to the MWC model, which assumes all-or-none activity within a cluster, Ising models allow each receptor to be independently active or inactive, based on its occupancy. The receptors interact cooperatively via equilibrium mechanisms (perhaps reflective of the lattice spatial structure present in real chemoreceptor complexes). A general Ising Hamiltonian for a chemoreceptor cluster is given by,
\begin{equation}
    H = \sum_{i=1}^N (E_s s_i + E_0 s_i \sigma_i - \mu \sigma_i) - \sum_{i\neq j } J_{ij} s_i s_j.
\end{equation}
The energy parameters have the same meanings as their MWC counterparts: $\mu = \log([L]/K_i)$ is the chemical potential (which depends on ligand concentration $[L]$ and dissociation constant $K_i$) and $E_s$ is the energy difference between active and inactive states, which is increased by $E_0$ when the receptor is occupied. Each receptor $s_i$ reacts individually to its occupancy $\sigma_i$, but the receptors interact cooperatively via the coupling $J_{ij}>0$.

Integrating out the binding degrees of freedom leads to a standard Ising model with an effective field,
\begin{equation}\label{effectiveIsingH}
    H_\mathrm{eff} = -\sum_{i \neq j} J_{ij} s_i s_j + h_\mathrm{eff} \sum_{i} s_i + N \log(1+e^\mu),
    \quad \quad h_\mathrm{eff} = E_s + \log \frac{1+e^\mu}{1+e^{\mu-E_0}}.
\end{equation}
We will consider the 1D (periodic) $J_{ij} = J (\delta_{i,j-1}+\delta_{i,j+1})/2$ and mean field $J_{ij} = J/(2N)$ Ising models. Given Eq.~(\ref{effectiveIsingH}), the partition function $Z = \sum_{\{s_i\}} \exp(-H_\mathrm{eff})$ is well known for each of these cases, from which the binding and activity can be readily computed $\expval{s} = -\dd{\log Z}/\dd{E_s}$ and $\expval{\sigma} = \dd{\log Z}/\dd{\mu}$. We omit the expressions here due to their complexity.

Given expressions for $\expval{s}$ and $\expval{\sigma}$, we can derive a parametric relation between the two $\expval{\sigma} = \tilde \sigma(\expval{s}, s_{\min}, s_{\max}|J, N)$, similar to that obtained in the main text for the MWC model. For the Ising models, $\expval{s}$ is generally not invertible, so we determine the inferred occupancy $\tilde \sigma$ numerically for a given receptor-receptor coupling $J$ and system size $N$. Fig.~\ref{fig:IsingTests} shows the parametric test for 1D and mean field Ising models applied to the Amin and Hazelbauer measurements~\cite{Amin2010Chemoreceptors}. For any choice of system size and receptor coupling, the data lie well off the diagonal, indicating these equilibrium models cannot simultaneously explain the binding and activity measurements.

\section{The nonequilibrium allosteric model captures binding and activity of other signaling proteins}

\subsection{\textit{E. coli} oxygen-sensing protein DosP}
\textit{E. coli} DosP is a c-di-GMP phosphodiesterase, whose enzymatic activity can be significantly enhanced by the binding of oxygen to DosP.
Previous experiments measured the ligand binding and phosphodiesterase activity at different oxygen concentrations and demonstrated that the results are incompatible with an equilibrium model~\cite{tuckerman_oxygen-sensing_2009}.
The same study proposed that the inconsistency with the equilibrium model could be attributed to ``memory effects"~\cite{tuckerman_oxygen-sensing_2009}, which suggests that the system operates out of equilibrium.
Here, we show that these measurements can be consistently explained under the framework of the nonequilibrium allosteric model presented in this work.

First, we carry out a parametric test to examine whether the measurements are consistent with an equilibrium MWC model with $N=4$ (DosP are known to form tetramers~\cite{yoshimura_relationships_2003}).
Due to the type of data available, the parametric test used here is slightly different from the one for the chemoreceptor measurements.
Specifically, we infer the mean ligand occupancy $\tilde{\sigma}$ from the mean activity $\expval{s}$ using the MWC model:
\begin{align}
    \expval{\sigma} = \tilde{\sigma}\qty(\expval{s}, t, [L]_{1/3}, [L]_{2/3}),
\end{align}
where $[L]_{1/3}$ and $[L]_{2/3}$ are the ligand concentration required to achieve $1/3$ or $2/3$ of normalized activity and $t=\frac{s_\mathrm{max}}{s_\mathrm{min}}$ is the ratio of maximum and minimum activities, which was measured to be 17~\cite{tuckerman_oxygen-sensing_2009}. The inferred occupancy $\tilde\sigma$ can be obtained by solving Eqs.~(2) and (3) of the main text. As shown by Fig.~6A of the main text, the parametric test found all data points deviating from the diagonal $\expval{\sigma}_\mathrm{exp}=\tilde\sigma$, revealing that the measurements are inconsistent with the equilibrium MWC model.

\begin{figure}
    \centering
    \includegraphics[width=0.8\linewidth]{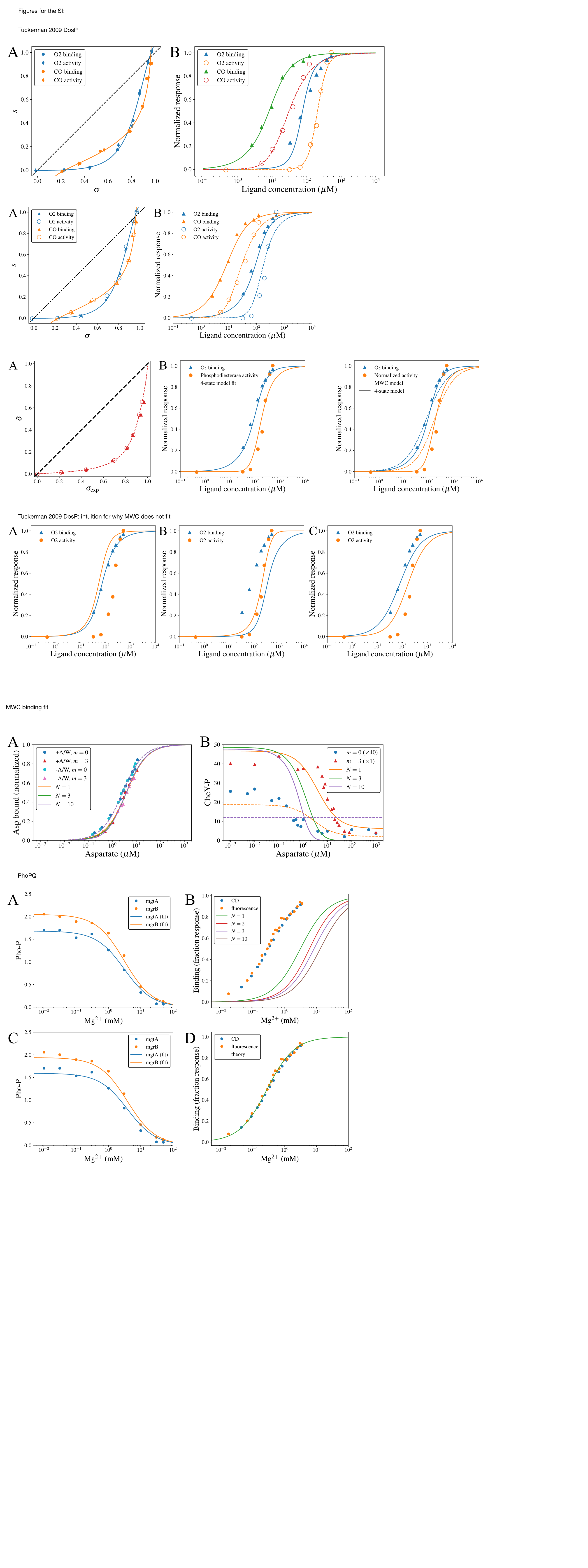}
    \caption{
    The MWC model cannot capture both the ligand binding and phosphodiesterase activity response of \textit{E. coli} DosP.
    (A) Fitting the binding curve results in significant inconsistency with the activity curve.
    (B) Fitting the activity curve results in inconsistency with the binding curve.
    (C) Fitting both curves simultaneously results in failure to capture their cooperativity (sharpness).
    For all fits, $N=4$ and $s_{\max}/s_{\min}=17$.
    }
    \label{fig:DosP MWC}
\end{figure}

The parametric test implies that the binding and activity data cannot be simultaneously explained under the framework of the MWC model, no matter how the fitting is done. To demonstrate this, we show representative fits of the MWC model to the data in Fig.~\ref{fig:DosP MWC}.
When fit to ligand binding (Fig.~\ref{fig:DosP MWC}A, blue line), the model incorrectly predicts the half-maximal concentration for activity (orange line).
Conversely, the model does not capture the half-maximal concentration for binding when fit to the activity curve (Fig.~\ref{fig:DosP MWC}B).
When fit to both curves, the model predicts the correct half-maximal concentrations but completely misses the sharpness of the curves, which describes cooperativity (Fig.~\ref{fig:DosP MWC}C). There exist other ways to fit the data, for example, by assigning different weights to activity and binding, respectively.
Nonetheless, none of them will be consistent with the measurements as demonstrated by the parametric test in Fig.~6A of the main text.

\begin{figure}
    \centering
    \includegraphics[width=0.5\linewidth]{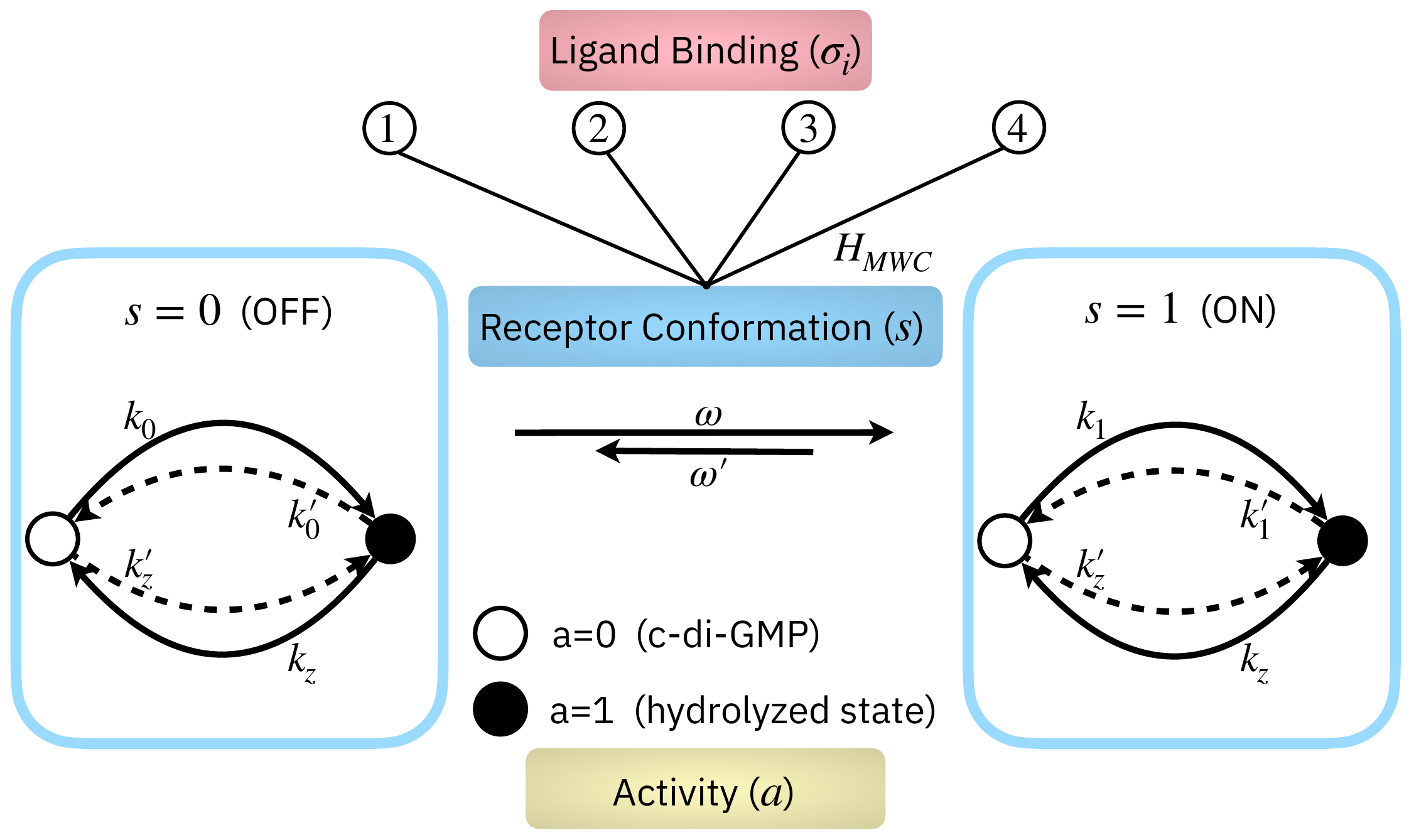}
    \caption{
        The nonequilibrium allosteric model with $k_0,k_0'>0$
        (namely, c-di-GMP hydrolysis is allowed both active and inactive states).
        For the DosP system, $a=0$ means c-di-GMP, and $a=1$ stands for its hydrolyzed state. $k_{0,1}$ are the rates of c-di-GMP hydrolysis, which leads to GMP through the intermediate pGpG. $k_z$ is the rate of c-di-GMP synthesis by diguanylate cyclases.
        The nonequilibrium driving is provided by GTP hydrolysis.
    }
    \label{fig:k0model}
\end{figure}

Next, we show that the nonequilibrium allosteric model is capable of simultaneously capturing binding and activity (Fig.~6B in the main text).
Here, we use a slightly generalized version of the model shown in the main text by allowing hydrolysis for both receptor conformation states $s=0$ and $s=1$,
whose rates are given by $k_0=k_ze^{E_{p0}}$ and $k_1=k_ze^{E_{p1}}$, respectively.
The model is shown in Fig.~\ref{fig:k0model}.
In the large dissipation limit, the mean activity is
\begin{align}
    \expval{a} = \frac{1}{1+\mathcal{P}},\quad \text{where\ }
    \mathcal{P} = \frac{\alpha_0+1}{\alpha_0 e^{E_{p1}} + e^{E_{p0}}},\
    \alpha_0 = e^{-E_s} \qty(\frac{1+[L]/K_i}{1+e^{-E_0} [L]/K_i})^{-N}.
\end{align}
The mean ligand occupancy is given by the solution to the MWC model,
\begin{align}
    \expval{\sigma} =\frac{[L]}{K_i+ [L]} (1 -\langle s \rangle) + \frac{ [L]}{e^{E_0}K_i+ [L]} \langle s \rangle.
\end{align}
where $\expval{s}=\frac{\alpha_0}{1+\alpha_0}$ is the mean receptor activity.
The model successfully fits the data as shown in Fig.~6B in the main text.

\bibliography{ref_chemotaxis}